\documentclass[twocolumn,showpacs,preprintnumbers]{revtex4}
\usepackage{amsmath}
\usepackage[dvips]{graphicx}
\usepackage{bm}

\begin{document}

\title{Angular dependence of coercivity in magnetic nanotubes}
\author{J. Escrig$^1$$^2$}
\author{M. Daub$^1$}
\author{P. Landeros$^3$}
\author{K. Nielsch$^1$$^4$}
\author{D. Altbir$^2$}
\affiliation{$^{1}$Max Planck Institute of Microstructure Physics,
Am Weinberg 2, 06120 Halle, Germany\\
$^{2}$Departamento de F\'{\i}sica, Universidad de Santiago de
Chile, USACH, Av. Ecuador 3493, Santiago, Chile\\
$^{3}$Departamento de F\'{i}sica, Universidad T\'{e}cnica Federico
Santa Mar\'{\i}a, Casilla 110-V, Valpara\'{\i}so,
Chile\\
$^{4}$Institute of Applied Physics, University of Hamburg,
Jungiusstr. 11, 20355 Hamburg, Germany}
\date{\today}

\begin{abstract}
The nucleation field for infinite magnetic nanotubes, in the case of a
magnetic field applied parallel to the long axis of the tubes, is calculated
as a function of their geometric parameters and compared with those produced
inside the pores of anodic alumina membranes by atomic layer deposition. We
also extended this result to the case of an angular dependence. We observed
a transition from curling-mode rotation to coherent-mode rotation as a
function of the angle in which the external magnetic field is applied.
Finally, we observed that the internal radii of the tubes favors the
magnetization curling reversal.
\end{abstract}

\maketitle

\section{Introduction}

Since the discovery of carbon nanotubes by Iijima in 1991, \cite{Iijima91}
intense attention has been paid to hollow tubular nanostructures because of
their particular significance for prospective applications. More recently,
magnetic nanotubes have been grown \cite{SSS+04, NCR+05, NCM+05, Wang05,
TGJ+06} motivating intense research in the field. Technological applications
of such systems require a deep knowledge and characterization of their
magnetic behavior. For example, changes in the internal radii are expected
to strongly affect the magnetization reversal mechanism \cite{LAE+07} and
thereby the overall magnetic response. \cite{ELA+07,SSS+04} The nature of
the magnetic tubes may be suitable for applications in biotechnology, where
magnetic nanostructures with low density, which can float in solutions, are
very desirable. \cite{NCR+05}

Coercivity is one of the most important properties of magnetic
materials for many present and future applications of permanent
magnets/magnetic materials, magnetic recording, and spin
electronics and, therefore, the understanding of magnetization
reversal mechanisms is a permanent challenge for researchers
involved in studying the properties of these materials. Recently,
Landeros \textit{et al}. \cite{LAE+07} found that, when a magnetic
field is applied parallel to the axis of a tube, the curling
reversal mode is the dominant magnetization reversal mechanism for
tubes with radii greater than $30$ nm. However, the angular
dependence of the coercivity in magnetic nanotubes has not been
studied yet, in spite of many works on this topic comprising
nanowires. \cite{WKF+99, ZHF+06} In this work we calculate the
angular dependence of the coercivity of Ni nanotubes, assuming
that the reversal of the magnetization occurs by means of one of
two possible modes: magnetization-curling mode ($V$) and
coherent-rotation mode ($C$).

Geometrically, tubes are characterized by their external and internal radii,
$R$ and $a$, respectively, and length $L$. It is convenient to define the
ratio $\beta \equiv a/R$, so that $\beta =0$ represents a solid cylinder and
$\beta $ close to $1$ corresponds to a very narrow tube. Besides, we
consider an external magnetic field $H_{a}$ applied in a direction defined
by $\theta _{0}$, with $\theta _{0}$ the angle of the applied field with the
tube axis, as illustrate in Fig. 1.
\begin{figure}[h]
\begin{center}
\includegraphics[width=7cm]{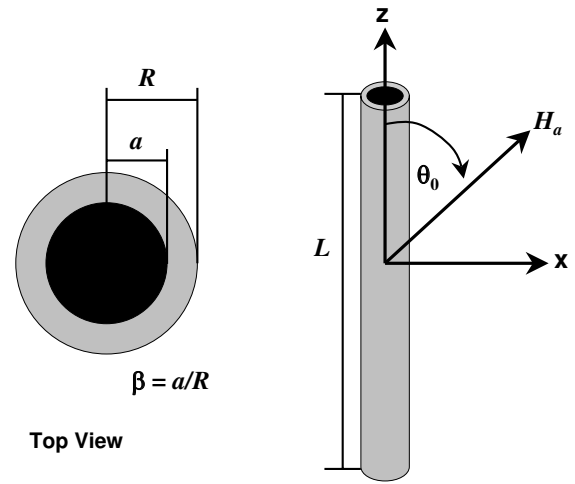}
\end{center}
\caption{Geometrical parameters of a magnetic nanotube.}
\end{figure}
In this paper we present an analytical model about the switching modes and
fields of infinite extended magnetic nanotubes in dependence of the
orientation of the magnetic field versus the nanotube axis. Additionally,
experimental data for the switching field of high-aspect ratio Ni nanotubes,
when the magnetic field is applied parallel to the tube axis, will be
compared with this micromagnetic model.

\section{Experimental details and results}

The high-aspect ratio Ni nanotubes were produced in porous
membranes with pore diameters of $180$, $220$ and $260$ nm. The
alumina membranes were coated by atomic layer deposition (ALD),
that consists of the sequential deposition of thin layers from two
different vapor-phase reactants, into a ferromagnetic Ni layer. We
used Nickeltocene (NiCp$_{2}$) and O$_{3}$ as reactants for the
deposition of a thin oxide layer, which was reduced after the ALD
process into a magnetic layer by annealing at $400$ $^{0}$C under
Ar + 5\% H$_{2}$ atmosphere. The deposition temperature was
between $270$ $^{0}$C and $330$ $^{0}$C with deposition rates of
$0.22-0.3$ \AA /cycle. Details about the preparation method can be
found elsewhere. \cite{DKG+07}

For SEM and TEM investigations, Ni nanotubes were deposited into
the alumina membrane between two layers of TiO$_{2}$. The
TiO$_{2}$ layers were also obtained by ALD \cite{KKW+06} and used
for adding a higher stability against oxidation to the nanotubes
and for preventing their damaging in the etching process. For TEM
measurements, the TiO$_{2}$/Ni/TiO$_{2}$ tubes were released by
etching the membrane in $1$ M NaOH and washing several times with
purified water. The magnetic properties of the Ni nanotubes were
measured by a superconducting quantum interference device (SQUID).
The Ni layer deposited on the top surface of the membrane was, in
all cases, removed by ion milling.
\begin{figure}[h]
\begin{center}
\includegraphics[width=6cm]{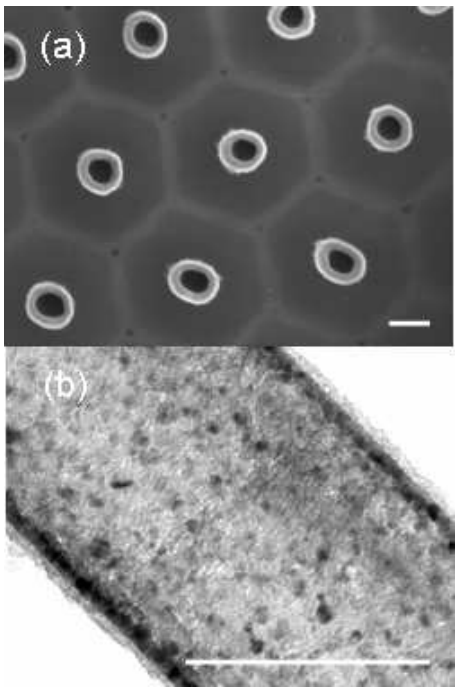}
\end{center}
\caption{SEM (a) and TEM (b) images for TiO$_{2}$/Ni/TiO$_{2}$ nanotubes of $%
180$ nm diameter and a Ni layer thickness of $11-12$ nm. The scale bar
indicates in both cases $150$ nm.}
\end{figure}

Figures 2(a) and (b) present typical SEM and TEM images for TiO$_{2}$/Ni/TiO$%
_{2}$ nanotubes with a diameter of around $180$ nm and a Ni layer thickness
of $11-12$ nm. Figure 3 shows the hysteresis cycles for three different
samples with diameters of $180$, $220$ and $260$ nm; pore length of around $%
5 $ $\mu $m and Ni layer thickness of $11-12$ nm. The measured
coercivities for these dimensions are, in all cases, around $200$
Oe ($1$ Oe $=10^{3}/4\pi $ Am$^{-1}$), higher than the coercivity
of bulk Ni (around $0.7$ Oe for Ni). \cite{Chikazumi64}
\begin{figure}[h]
\begin{center}
\includegraphics[width=7cm]{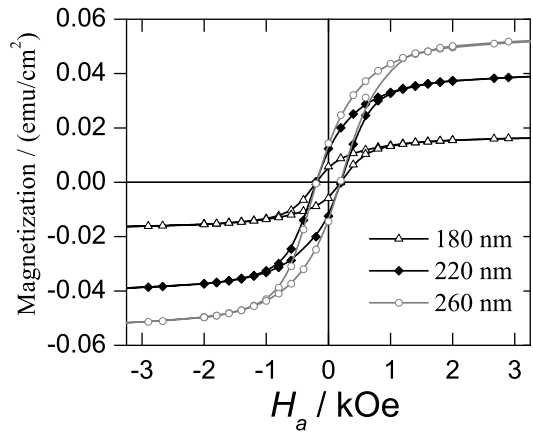}
\end{center}
\caption{Hysteresis cycles for Ni nanotubes as a function of the
tube diameter. The magnetic field is applied parallel to the tube
axis. The Ni layer thickness is the same in all cases ($11-12$
nm).}
\end{figure}

\section{Model and discussion}

\subsection{Magnetic field applied parallel to the long axis}

As pointed out by Landeros \textit{et al}. \cite{LAE+07}, the
curling reversal mode is the dominant magnetization reversal
process in magnetic nanotubes. The magnetization curling mode was
proposed by Frei \textit{et al}. \cite{FST57} and has been used to
investigate the magnetic switching of films \cite{IHS+97} and
particles with different geometries, like spheres, \cite{FST57}
prolate ellipsoids \cite{FST57,ST63} and cylinders. \cite{IS89}
However, for simplicity, expressions for the nucleation field
obtained using infinite cylinders are used.

In the case of a magnetic field applied parallel to the long axis
of an infinite tube, we present an analytical approach to the
nucleation field obtained from a Ritz model. Calculations are
shown in the Appendix and lead us to write%
\begin{equation}
\frac{H_{n}^{V}}{M_{0}}=-\frac{2K}{\mu _{0}M_{0}^{2}}-\alpha \left( \beta
\right) \frac{L_{ex}^{2}}{R^{2}}\ ,  \label{1}
\end{equation}%
where $L_{x}=\sqrt{2A/\mu _{0}M_{0}^{2}}$, $M_{0}$ is the saturation
magnetization, $A$ is the stiffness constant of the magnetic material, $K$
is the magnetocrystalline anisotropy constant and
\begin{equation}
\alpha \left( \beta \right) =\frac{8}{3}\frac{\left( 14-13\beta ^{2}+5\beta
^{4}\right) }{\left( 11+11\beta ^{2}-7\beta ^{4}+\beta ^{6}\right) }\ .
\label{2}
\end{equation}%
Equation (\ref{1}) has been previously obtained by Chang \textit{et al}.
\cite{CLY94} starting from the Brown's equations. They obtained $\alpha
(\beta )\equiv q^{2}$, with $q$ satisfying
\begin{equation}
\frac{qJ_{0}(q)-J_{1}(q)}{qY_{0}(q)-Y_{1}(q)}-\frac{\beta qJ_{0}(\beta
q)-J_{1}(\beta q)}{\beta qY_{0}(\beta q)-Y_{1}(\beta q)}=0\ .  \label{3}
\end{equation}%
Here $J_{p}\left( z\right) $ and $Y_{p}\left( z\right) $ are Bessel
functions of the first and second kind, respectively. Equation (\ref{3}) has
an infinite number of solutions, out of which only the one with the smallest
nucleation field has to be considered. \cite{Aharoni96} Therefore, the
nucleation field depends on $\alpha (\beta )$, which is related to the
internal and external radii of the tube. Figure 4 illustrate $\alpha $ as a
function of $\beta $, obtained numerically from Brown's equations, Eq. (\ref%
{3}), and by means of Eq. (\ref{2}) using our analytical approach (Ritz
method). We observe that both results are similar, showing a perfect
agreement for nanowires $\left( \beta =0\right) $ and very narrow nanotubes $%
\left( \beta \rightarrow 1\right) $. This behavior lead us to simply use the
analytical expression, Eq. (\ref{2}), to calculate the nucleation field of a
magnetic nanotube.
\begin{figure}[h]
\begin{center}
\includegraphics[width=7cm]{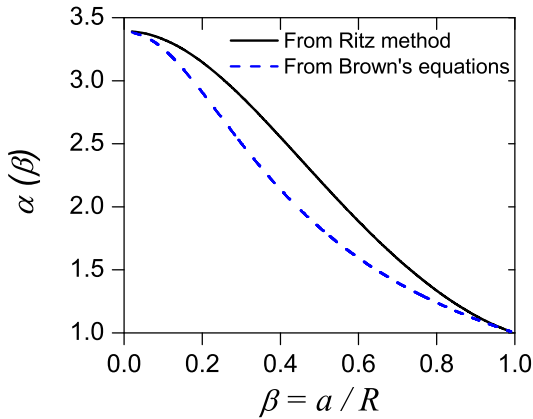}
\end{center}
\caption{$\protect\alpha $ as a function of $\protect\beta $ from Eq. (%
\protect\ref{3}) (dashed line) and by means of our analytical approach, Eq. (%
\protect\ref{2}) (solid line).}
\end{figure}
Therefore, $\alpha (\beta )$ is a decreasing function of the aspect ratio $%
\beta =a/R$ of the tube. It is clear that, for $\beta =0$, i.e.,
for an infinite cylinder (or nanowire), $\alpha \left( 0\right)
=1.08\pi $, as previously calculated by Shtrikman \textit{et al}.
\cite{ST63}

As pointed by Aharoni, \cite{Aharoni97} for a prolate spheroid
with $\theta _{0}=0$, a jump of the magnetization at, or near, the
curling nucleation field occurs. Therefore, the coercivity is
quite close to the absolute value of the nucleation field. Then,
we assumed here that $-H_{n}^{V}$ is a good approximation to the
coercivity, $H_{c}^{V}$, when the reversal is by
curling, as in other works considering infinite cylinders. \cite%
{Ishii91,ST63}

Now we investigate the validity of Eqs. (\ref{1}) and (\ref{2}) by
calculating the coercivity for different samples experimentally
investigated. Table I summarizes the geometrical parameters of the arrays,
measured $H_{c}$, and calculated $H_{c}^{V}$. In our calculations we used $%
M_{0}=4.85\times 10^{5}$ A/m and $K=4.5\times 10^{3}$ J/m$^{3}$,
both taken from Ref. [22] at room temperature. In the same
reference it is pointed that $A$ ranges for any material from $1$
to $2\times 10^{-11}$ J/m. However, by means of field-dependent
elastic small-angle neutron scattering (SANS), the
exchange-stiffness constant $A$ for Ni was determined by Michels
\textit{et al}. \cite{MWW+00} At ambient temperature $A=7.6\pm
0.3\times 10^{-12}$ J/m was reported for nanocrystalline samples
while $9.2\pm 0.2\times 10^{-12}$ J/m was obtained at $T=5$ K. For
our calculations we choose $A=7.6\times 10^{-12}$ J/m.
\begin{table}[h]
\caption{Parameters for different Ni nanotube arrays with
different tube diameters ($2R$) and $5$ $\protect\mu $m tube
length. $H_{c}$ represents the experimentally measured coercive
field. }
\begin{center}
\begin{tabular}{|c|c|c|c|}
\hline
$2R\left( nm\right) $ & $\beta $ & $H_{c}\left( Oe\right) $ & $%
H_{c}^{V}\left( Oe\right) $ \\ \hline\hline
$180$ & $0.88$ & $200$ & $231$ \\ \hline
$220$ & $0.90$ & $200$ & $215$ \\ \hline
$260$ & $0.91$ & $200$ & $206$ \\ \hline
\end{tabular}%
\end{center}
\end{table}

In the measured samples, small variations of around $5$ Oe for the
coercive field were detected. However, we consider these
variations to be within the range of measurement error. From the
work of AlMawlawi {\it et al} \cite{ACM91} small variations of the
coercivity are observed for nanowires as a function of the aspect
ratio for aspect ratios higher than 20. In order to observe small
variations in the coercivity, very well geometrically
characterized samples need to be measured with well controlled
inter-element interactions. In our samples the center to center
distance is fixed, and then the strength of interactions is also
different from one sample to other, making a direct comparison
difficult. The computed values for the coercivity are larger than
the experimental data. We ascribe such differences between
calculations and experimental results to the interaction of each
tube with the stray field produce by the array. This field
originated in the effective antiferromagnetic coupling between
neighboring tubes, which reduces the coercive field as previously
demonstrated in nanowires. \cite{Hertel01, BAA+06} As $\beta
\rightarrow 1$ the contribution of the stray field diminishes and
a better agreement between theory and experiment is obtained.
Besides, a fully realistic approach needs to consider a finite
nanotube, making much more complex the calculations and the
expression for the energy. Therefore, the small discrepancy
between experiments and model can be regarded as the result of our
models simplification.

\subsection{Angular dependence of the coercivity}

We now proceed to investigate the angular dependence of the
coercivity for magnetic nanotubes. We calculate the coercive field
$H_{c}^{k}$ assuming each of the previously mentioned reversal
mechanisms, $k=C$ (coherent) or $V$ (curling).

\textbf{Coherent-mode rotation (C)}

The angular dependence of the nucleation for a coherent magnetization
reversal was calculated by Stoner-Wohlfarth \cite{SW48} and gives%
\begin{equation}
\frac{H_{n}^{C}}{M_{0}}=-\frac{1-3N_{z}}{2}\frac{\sqrt{1-t^{2}+t^{4}}}{%
1+t^{2}}\ ,  \label{4}
\end{equation}%
where $t=\tan ^{\frac{1}{3}}\left( \theta _{0}\right) $ and $N_{z}$
corresponds to the demagnetizing factor of the ellipsoid along $z$. For a
tube, the demagnetizing factor can be calculated from Escrig \textit{et al}.
\cite{ELA+07}, so that
\[
N_{z}=\frac{2R}{L\left( 1-\beta ^{2}\right) }\int_{0}^{\infty }\frac{dq}{%
q^{2}}\left( J_{1}\left( q\right) -\beta J_{1}\left( \beta q\right) \right)
^{2}\left( 1-e^{-q\frac{L}{R}}\right) \ .
\]%
In the Stoner-Wolhfarth model \cite{SW48} the nucleation field,
$H_{n}^{C}$, does not represents the coercivity, $H_{c}^{C}$, in
all cases. However,
from the discussion on p 21 [27], the coercivity can be written as%
\[
H_{c}^{C}=\left\{
\begin{array}{c}
\left\vert H_{n}^{C}\right\vert \quad 0\leq \theta _{0}\leq \pi /4 \\
2\left\vert H_{n}^{C}\left( \theta _{0}=\pi /4\right) \right\vert
-\left\vert H_{n}^{C}\right\vert \quad \pi /4\leq \theta _{0}\leq \pi /2%
\end{array}%
\right. \ .
\]

\textbf{Curling-mode rotation (V)}

The angular dependence of the curling nucleation field in a finite
prolate spheroid was obtained by Aharoni. \cite{Aharoni97} By
extending the expression for the switching field to take into
account the internal radii of
tubes, we obtain%
\begin{equation}
\frac{H_{n}^{V}}{M_{0}}=\frac{\left( N_{z}-\frac{\alpha \left( \beta \right)
L_{x}^{2}}{R^{2}}\right) \left( N_{x}-\frac{\alpha \left( \beta \right)
L_{x}^{2}}{R^{2}}\right) }{\sqrt{\left( N_{z}-\frac{\alpha \left( \beta
\right) L_{x}^{2}}{R^{2}}\right) ^{2}\sin ^{2}\theta _{0}+\left( N_{x}-\frac{%
\alpha \left( \beta \right) L_{x}^{2}}{R^{2}}\right) ^{2}\cos ^{2}\theta _{0}%
}}  \label{5}
\end{equation}

Figure 5 illustrates the coercive field as a function of $\theta
_{0}$ for an infinite nickel nanotube. Dashed curves represent the
coercivity of a nanotube due to a reversal curling mode. The
cutoff of the curves corresponds to the transition angles, $\theta
_{0}^{T}$, at which a coherent reversal mode appears. The most
remarkable feature of these curves is that the general shape for
nanotubes is similar to the one for nanowires. Differences, of
course, are far from being negligible, and the internal radii must
be taken into account in any proper analysis of experimental data.
For example, using Eq. (\ref{5}) we found that for a very thin
nanotube ($\beta =0.9$) with $R=50$ nm, the coercivity is almost
the same as in a nanowire with $R=100$ nm. We also observe in this
figure that, for $ \theta _{0}>60^{0}$, a small uncertainly in the
measurement of $\theta _{0}$ can cause large changes in the
coercivity.
\begin{figure}[h]
\begin{center}
\includegraphics[width=7cm]{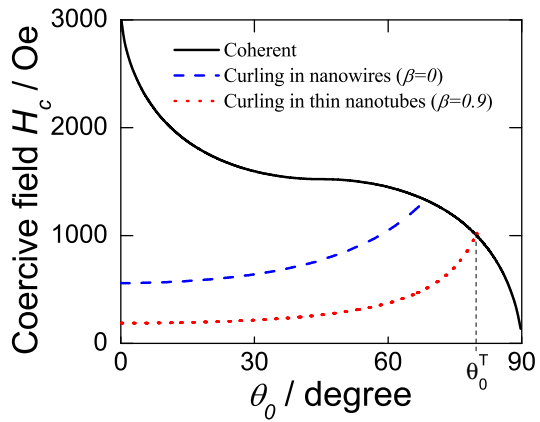}
\end{center}
\caption{Angular dependence of the coercivity, $H_{c}$, in an infinite
nickel nanotube with $R=50$ nm. The solid line represents the coherent
rotation, the dashed line corresponds to the curling mode for $\protect\beta %
=0$, and the dotted line illustrates our results for the curling mode in a
tube defined by $\protect\beta =0.9$.}
\end{figure}

We now investigate the dependence of the coercivity as a function
of $R$. We illustrate our results with trajectories of the
transition angle, $\theta _{0}^{T}$, in Fig. 6 for Ni. Each line
separates the coherent reversal mode (upper) from the curling
reversal mode (lower). Results for nickel also represent iron
oxide tubes because of the similar magnetic parameters of both
materials. In the considered range of parameters, we observe that
an increase of the external radii, $R$, or $\beta $ (see Fig. 7)
results in an increase of the transition angle, $\theta _{0}^{T}$,
enhancing the region of stability of the curling reversal mode.
However, the dependence of the coercivity on $R$ is stronger than
on $\beta $.
\begin{figure}[h]
\begin{center}
\includegraphics[width=7cm]{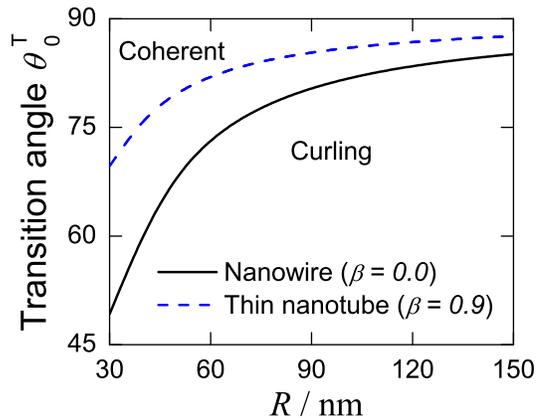}
\end{center}
\caption{Trajectories of the transition angle, $\protect\theta
_{0}^{T}$, as a function of $R$, for $\protect\beta =0$ (solid
line) and $\protect\beta =0.9$ (dashed line). }
\end{figure}
\begin{figure}[h]
\begin{center}
\includegraphics[width=7cm]{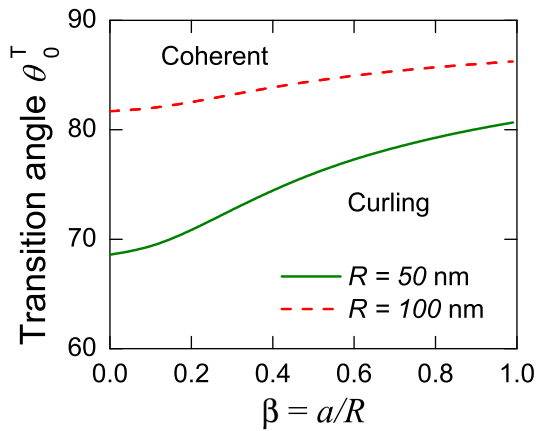}
\end{center}
\caption{Trajectories of the transition angle, $\protect\theta _{0}^{T}$, as
a function of $\protect\beta $, for $R=50$ nm (solid line) and $R=100$ nm
(dashed line). }
\end{figure}

\section{Conclusions}

In conclusion, by means of theoretical studies and experimental
measurements, we have investigated the coercivity in magnetic
nanotubes. We have obtained Ni nanotubes by atomic layer
deposition into alumina membranes. ALD proves to be a powerful
technique, which allows us to have a very precise control of layer
growth. We have also derived an analytical expression that allows
one to obtain the coercivity when a magnetic field is applied
parallel to the tube axis. Good agreement between the measured
magnetic properties of Ni nanotubes and theoretical calculations
is obtained. Finally, this calculation has been extended to the
case of an angular dependence of the coercivity, where a
transition from curling-mode rotation to coherent-mode rotation
has been observed. However, further experimental work remains to
be done in order to observe this transition.

\begin{acknowledgments}
This work has been partially supported by the German Federal Ministry
for Education and Research (BMBF, project No 03N8701) in Germany,
and Millenium Science Nucleus "Basic and Applied Magnetism"
P06-022F in Chile. PBCT (PSD-031) and AFOSR (Award N$^{o}$
FA9550-07-1-0040) are also acknowledged.
\end{acknowledgments}

\appendix

\section*{Appendix: nucleation field for an infinite nanotube from a Ritz
model}

\setcounter{section}{1}.

We use the term \textit{curling mode} here not in reference to an
eigenfunction of Brown's equation. In our case we replace the spatial
dependence of the curling eigenfunction by a Ritz model that approximates
the curling eigenmode, which turns out to be quite simple. We use the Ritz
model previously used by Ishii \textit{et al.} for infinite cylinders. \cite%
{Ishii91} We assume the following model for the magnetization%
\begin{eqnarray*}
\varepsilon m_{x} &=&-\varepsilon \left( \frac{r}{R}-\frac{r^{3}}{3R^{3}}%
\right) \sin \phi \\
\varepsilon m_{y} &=&\varepsilon \left( \frac{r}{R}-\frac{r^{3}}{3R^{3}}%
\right) \cos \phi \\
\varepsilon m_{z} &=&1-\frac{\varepsilon ^{2}}{2}\left(
m_{x}^{2}+m_{y}^{2}\right) \ ,
\end{eqnarray*}%
which satisfies $\varepsilon ^{2}m_{x}^{2}+\varepsilon
^{2}m_{y}^{2}+\varepsilon ^{2}m_{z}^{2}=1+\vartheta \left( \varepsilon
^{4}\right) $, with $\varepsilon $ an infinitesimal parameter.

The total energy density $E$ is generally given by the sum of four terms
corresponding to the magnetostatic $E_{dip}^{V}$, the exchange $E_{ex}^{V}$,
the magnetocrystalline anisotropy $E_{K}^{V}$, and the Zeeman $E_{h}^{V}$
(resulting from the interaction between $\mathbf{M}$ and an external field $%
\mathbf{H}_{a}$), can be calculated using the well known continuum theory of
ferromagnetism. \cite{Aharoni96} Because in this case we do not have charges
in the surface, the contribution from the magnetostatic energy density
results equal to zero, $E_{dip}^{V}=0$. The exchange energy density is given
by $E_{ex}^{V}=\left( A/s\right) \int \sum \left( \mathbf{\nabla }%
\varepsilon m_{i}\right) ^{2}dv$, with $s=\pi R^{2}\left( 1-\beta
^{2}\right) $ and $m_{i}=M_{i}/M_{0}$ $\left( i=x,y,z\right) $. Thus, we
obtain
\[
E_{ex}^{V}=\varepsilon ^{2}\frac{\mu _{0}M_{0}^{2}}{27}\left( \frac{L_{x}}{R}%
\right) ^{2}\left( 14-13\beta ^{2}+5\beta ^{4}\right) \ ,
\]%
The magnetocrystalline anisotropy energy density can be written as $%
E_{K}^{V}=\left( K/s\right) \int \left( \varepsilon
^{2}m_{x}^{2}+\varepsilon ^{2}m_{y}^{2}\right) dv$, so that
\[
E_{K}^{V}=\varepsilon ^{2}\frac{K}{36}\left( 11+11\beta ^{2}-7\beta
^{4}+\beta ^{6}\right) \ .
\]%
Finally we consider the Zeeman energy density, which is given by $%
E_{h}^{V}=-\left( \mu _{0}H_{a}M_{0}/s\right) \int \varepsilon m_{z}dv$.
Thus, we obtain%
\[
E_{h}^{V}=-\mu _{0}H_{a}M_{0}\left[ 1-\frac{\varepsilon ^{2}}{72}\left(
11+11\beta ^{2}-7\beta ^{4}+\beta ^{6}\right) \right] \ .
\]%
Now we are able to obtain the total energy density $E^{V}$. The second
variation in the magnetic energy density $E^{V}$ with respect to a small
deviation $\varepsilon \mathbf{m}$ from $\mathbf{\hat{z}}$, where $\mathbf{%
\hat{z}}$ is the unit vector along the magnetization, must be positive at
the equilibrium state and zero at nucleation. Therefore,
\[
\frac{H_{n}^{V}}{M_{0}}=-\frac{2K}{\mu _{0}M_{0}^{2}}-\alpha \left( \beta
\right) \left( \frac{L_{x}}{R}\right) ^{2}\ ,
\]%
where
\[
\alpha \left( \beta \right) =\frac{8}{3}\frac{\left( 14-13\beta ^{2}+5\beta
^{4}\right) }{\left( 11+11\beta ^{2}-7\beta ^{4}+\beta ^{6}\right) }\ .
\]

\end{document}